\documentclass[showpacs,preprintnumbers,twocolumn,prb]{revtex4}%
\usepackage{amssymb}
\usepackage{amsfonts}
\usepackage{amsmath}
\usepackage{graphicx}
\usepackage{dcolumn}
\usepackage{bm}%
\begin{document}
\title{Signal amplification in a nanomechanical Duffing resonator via stochastic resonance}
\author{Ronen Almog}
\author{Stav Zaitsev}
\author{Oleg Shtempluck }
\author{Eyal Buks}
\affiliation{Department of Electrical Engineering, Technion, Haifa 32000 Israel}

\begin{abstract}
We experimentally study stochastic resonance in a nonlinear bistable
nanomechanical resonator. The device consists of a PdAu doubly clamped beam
serving as a nanomechanical resonator excited capacitively by an adjacent gate
electrode and its vibrations are detected optically. The resonator is tuned to
its bistability region by an intense pump near a point of equal transition
rates between its two metastable states. The pump is amplitude modulated,
inducing modulation of the activation barrier between the states. When noise
is added to the excitation, the resonator's displacement exhibits noise
dependent amplification of the modulation signal. We measure the resonator's
response in the time and frequency domains, the spectral amplification and the
statistical distribution of the jump time.

\end{abstract}
\maketitle

Stochastic resonance (SR) is a phenomenon in which an appropriate amount of
noise is used to amplify a periodic signal acting on a bistable nonlinear
system.\cite{Benzi1}$^{-}$\cite{ShatokhinRPP} SR has been demonstrated
experimentally in electrical, optical, superconducting, and neuronal
systems.\cite{Fauve}$^{-}$\cite{Levin} SR could be used for amplification in
nanomechanical devices in order to improve force detection
sensitivity.\cite{Badzey}$^{-}$\cite{Chan} Nanomechanical resonators operating
in their nonlinear regime exhibit the well known Duffing bistability. In a
Duffing oscillator\cite{Nayfebook}, above a critical excitation amplitude, the
response becomes a multi-valued function of the frequency in some finite
frequency range, and the system becomes bistable (with a low amplitude state
$S_{l}$ and a high amplitude one $S_{h})$ with jump points in the frequency
response. In the presence of noise, the oscillator can occasionally overcome
the activation barrier and hop between the states. \cite{Dykman} When an
oscillator is excited in the bistability region near a point of equal
transition rates between its states, an amplitude modulation (AM) of the force
could be amplified by noise when the noise dependent transition rate is
comparable to twice the modulation frequency. This type of SR, where the
bistability property depends on the driving force, is usually referred to as
\textit{high frequency SR}.\cite{Chan}$^{,}$\cite{DykmanSNSP}

In this paper we demonstrate high frequency SR in our nanomechanical resonator
and measure the noise dependent amplification. Our study extends previous
work\cite{Badzey}$^{-}$\cite{Chan} by characterizing the SR by spectral
amplification\cite{Jung}, and by measuring the statistical distribution of the
jump time at SR condition. The system under study consists of a nonlinear
doubly clamped nanomechanical PdAu beam, excited capacitively by an adjacent
gate electrode.\cite{Almog1}$^{-}$\cite{Almog2} The device is shown in the
inset of Fig. 2. The bistability region of the device is found by exciting the
resonator with a harmonic pump signal, sweeping its amplitude upward and then
downward for constant pump frequency, calculating the difference between the
two responses, and repeating for a range of frequencies. The result is shown
in Fig. 1a. An example of a pump amplitude hysteresis loop for a constant pump
frequency of 520.58$%
\operatorname{kHz}%
$ (the broken line in Fig. 1a) is shown in Fig. 1b. When the pump is amplitude
modulated without additional noise, the resonator will respond with small
amplitude oscillation in the respective hysteresis branch (vertical black line
in Fig. 1b). To bring the system into SR, the resonator is tuned to its
bistability region by an intense pump near a point of equal transition rates
between its states. Next, the pump is amplitude modulated, inducing thus
modulation of the activation barrier between the states and modulating the
transition rates $\Gamma_{1}$ and $\Gamma_{2}$ of the noise-driven transitions
$S_{l}\rightarrow S_{h}$ and $S_{h}\rightarrow S_{l}$ respectively.
\begin{figure}
[h]
\begin{center}
\includegraphics[
height=4.8551in,
width=3.4342in
]%
{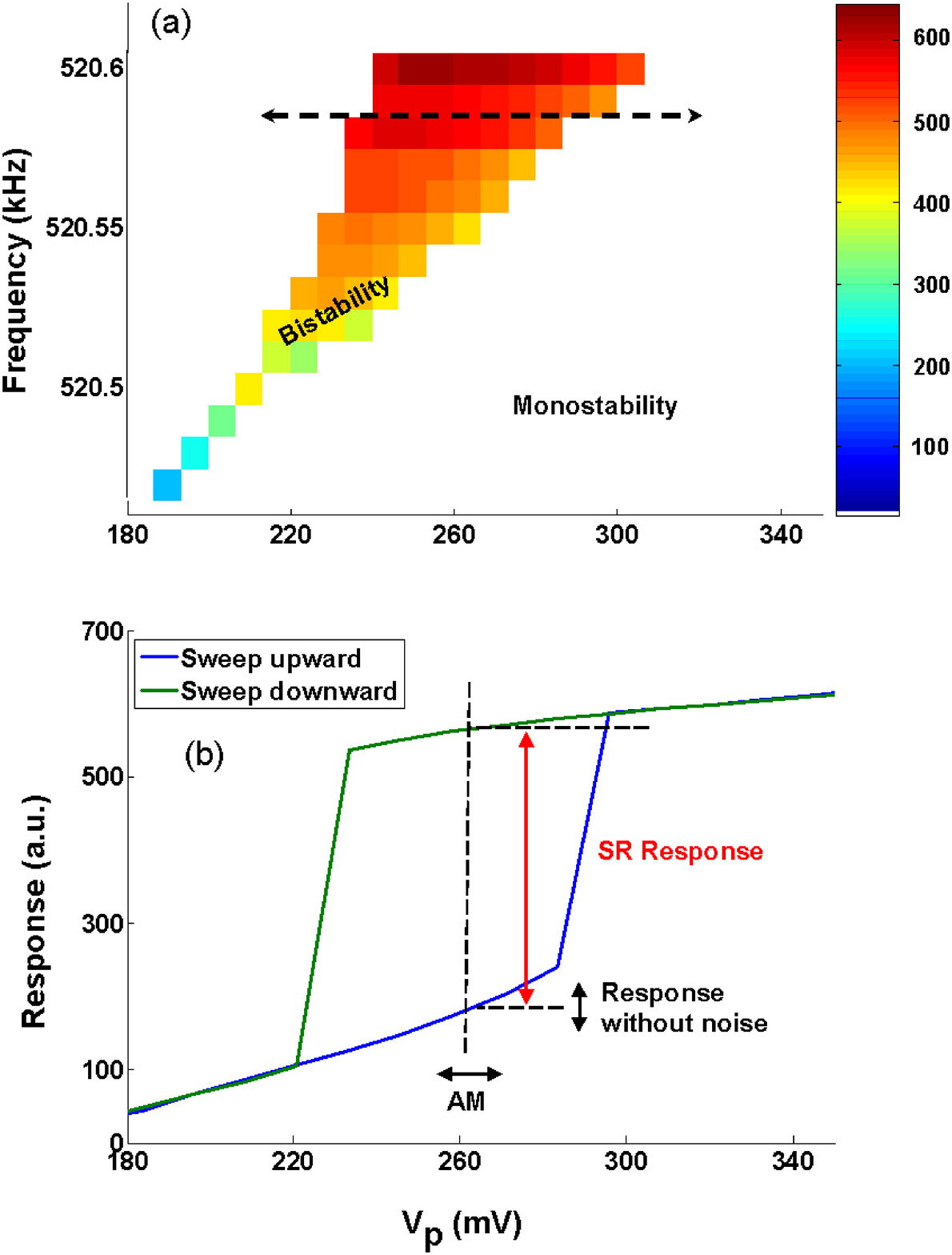}%
\caption{(Color online) (a) Measurement of the bistability region. (b) Pump
amplitude hysteresis loop for a constant pump frequency of
520.58$\operatorname{kHz}$ (the broken line in Fig. 1a). The vertical arrows
show the response to a small AM (horizontal arrow).}%
\end{center}
\end{figure}
When an appropriate amount of noise is added, the resonator will hop from one
state to the other in synchronization with the modulation signal and with
large amplitude (vertical red line in Fig. 1b). The working point (pump
amplitude and frequency) is determined such that $\Gamma_{1}\simeq$
$\Gamma_{2}$.

The nonlinear dynamics of the fundamental mode of a doubly clamped beam
excited by an external force per unit mass $F(t)$ can be described by a
Duffing oscillator equation for a single degree of freedom $x$
\begin{equation}
\ddot{x}+2\mu\dot{x}+\omega_{0}^{2}(1+\kappa x^{2})x=F(t),
\end{equation}
where $\mu$ is the damping constant, $\omega_{0}/2\pi$ is the resonance
frequency of the fundamental mode of the oscillator, and $\kappa$ is the cubic
nonlinear constant. Our resonator has a quality factor $Q=\omega_{0}%
/2\mu\approx2000$ (at $10^{-5}%
\operatorname{torr}%
)$ and its fundamental mode resonance frequency is $\omega_{0}/2\pi\simeq
$520.4$%
\operatorname{kHz}%
.$ The resonator is excited by an applied force $F(t)=f_{p}%
(1+A_{\operatorname{mod}}\cos\Omega t)\cos(\omega_{p}t)+F_{n}(t)$ composed of
an amplitude modulated pump signal with amplitude $f_{p},$ frequency
$\omega_{p},$ modulation frequency $\Omega\ll\omega_{p}$ and modulation depth
$A_{\operatorname{mod}}<1$, and $F_{\mathrm{n}}(t)$ is a zero-mean Gaussian
white noise with autocorrelation function $\left\langle F_{\mathrm{n}%
}(t)F_{\mathrm{n}}(0)\right\rangle =2D\delta(t)$ and noise intensity $D$. This
is achieved by applying a voltage of the form $V=V_{dc}+V_{p}%
(1+A_{\operatorname{mod}}\cos\Omega t)\cos(\omega_{p}t)+V_{n}(t)$ where
$V_{dc}$ is a DC bias (employed for tuning the resonance frequency), $V_{p}$
is the pump amplitude, and $V_{n}(t)$ is the applied voltage noise. The
voltage noise intensity is $I\equiv\left\langle V_{n}^{2}(t)\right\rangle
^{1/2}$ and $V_{p},I\ll V_{dc}$.

The displacement spectral density can be expressed as
\begin{equation}
S_{x}(\omega)=%
{\displaystyle\sum\limits_{k=-\infty}^{k=\infty}}
A_{k}(D)\delta(\omega_{p}+k\Omega)+S_{nx}(\omega),
\end{equation}
composed of delta peaks at the mixing products $\omega_{p}+k\Omega,$
$k=0,\pm1,\pm2...$, with noise dependent amplitudes $A_{k}(D)$, and a
background spectral density of the noise denoted by $S_{nx}(\omega).$ In order
to characterize the noise dependent amplification, we define a spectral
amplification parameter $\eta_{k}$ by%
\begin{equation}
\eta_{k}(D)=A_{k}(D)/A_{k}(D=0).
\end{equation}

A schematic diagram of the experimental setup employed for measuring SR is
depicted in Fig. 2.%
\begin{figure}
[h]
\begin{center}
\includegraphics[
height=2.2866in,
width=3.4307in
]%
{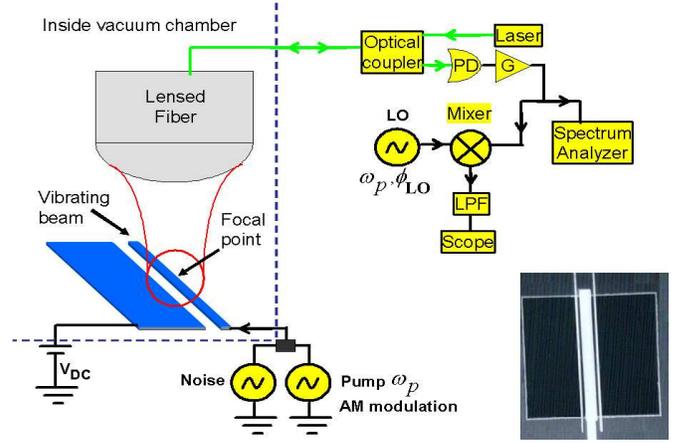}%
\caption{(Color online) The experimental setup. The device consists of a
suspended doubly clamped nanomechanical resonator. The resonator is excited by
two arbitrary waveform generators (one is used for the pump and the second for
the noise). The resonator's vibrations are detected optically. The inset shows
an electron micrograph of the device.}%
\end{center}
\end{figure}
The resonator is excited by two sources (pump and noise) and its vibrations
are detected optically using a knife-edge technique. The device is located at
the focal point of a lensed fiber which is used to focus laser light at the
beam and to collect the reflected light back to the fiber and to a
photodetector (PD). To measure the response, the PD signal is amplified, mixed
with a local oscillator (LO), and low pass filtered. The spectrum around
$\omega_{p}$ of the amplified PD signal is measured using a spectrum analyzer.
The measurement is done in vacuum ($10^{-5}$ $%
\operatorname{torr}%
$) and at room temperature. The resonator length is $100%
\operatorname{\mu m}%
$, width $600%
\operatorname{nm}%
$, and thickness $250%
\operatorname{nm}%
$. The gap separating the doubly clamped beam and the stationary side
electrode is $4%
\operatorname{\mu m}%
$ wide. The device is fabricated using bulk nano-machining process together
with electron beam lithography \cite{ebmlr}.

Typical results of SR measured in the time and frequency domains are shown at
the left and right sides respectively of Fig. 3 for five voltage noise
intensities (panels (a)-(e))$.$ Here $\Omega$=20$%
\operatorname{Hz}%
$, $A_{\operatorname{mod}}=10\%,$ and $V_{dc}=25%
\operatorname{V}%
.$ The blue dotted line drawn in the time domain represents the modulation
signal.
\begin{figure}
[h]
\begin{center}
\includegraphics[
height=2.3436in,
width=3.4074in
]%
{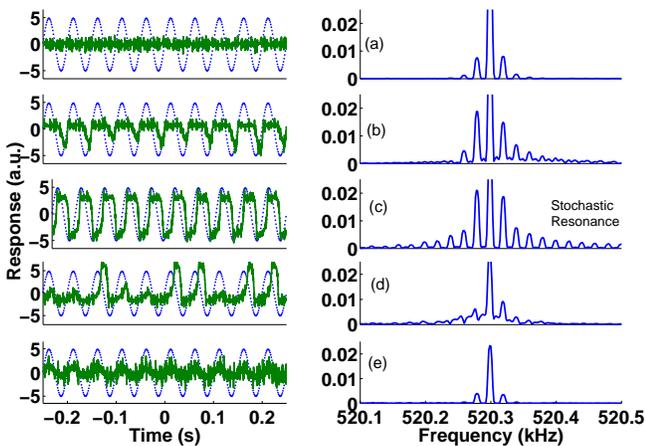}%
\caption{(Color online). Panels $(a)-(e)$ exhibit typical snapshots of the
resonator's response in the time domain (left) and in the frequency domain
(right) as the input voltage noise intensity is increased. The dotted line in
the time domain represents the modulation signal.}%
\label{timeSpecttraces}%
\end{center}
\end{figure}
The voltage noise intensities (a) $1%
\operatorname{mV}%
$ and (b) $349%
\operatorname{mV}%
$ correspond to low noise levels below the value corresponding to SR . Panel
(a) shows response without jumps. Panel (b) shows a response containing few
arbitrary jumps. The voltage noise intensity (c) $464%
\operatorname{mV}%
$ corresponds to SR condition where every half cycle, the resonator jumps to
the other metastable state. The voltage noise intensities (d) $530%
\operatorname{mV}%
$ and (e) $600%
\operatorname{mV}%
$ are higher than the the value corresponding to SR. In panel (d)$,$ as in
Ref.,\cite{Badzey} the resonator stays in the $S_{l}$ state with few jumps to
the $S_{h}$ state. In panel (e), the high noise almost completely screens the
signal. In the frequency domain displayed at the right side of Fig.
\ref{timeSpecttraces}$,$ the fundamental frequency and the mixing products can
be seen. At SR, the spectrum contains high order mixing products.

The dependence of the spectral amplification $\eta_{k}(D)$ ($k=\pm1$ and
$k=\pm3)$ on voltage noise intensity $I$ is shown in Fig. 4a and Fig. 4b
respectively. Here $\Omega=30%
\operatorname{Hz}%
$, $A_{\operatorname{mod}}=10\%,$ and the optimal noise intensity for maximal
amplification $I=750%
\operatorname{mV}%
.$ The amplification $\eta_{\pm1}$have maximal value of $5$ whereas $\eta
_{\pm3} $ have maximal value of $40$.%
\begin{figure}
[t]
\begin{center}
\includegraphics[
height=2.38in,
width=3.4592in
]%
{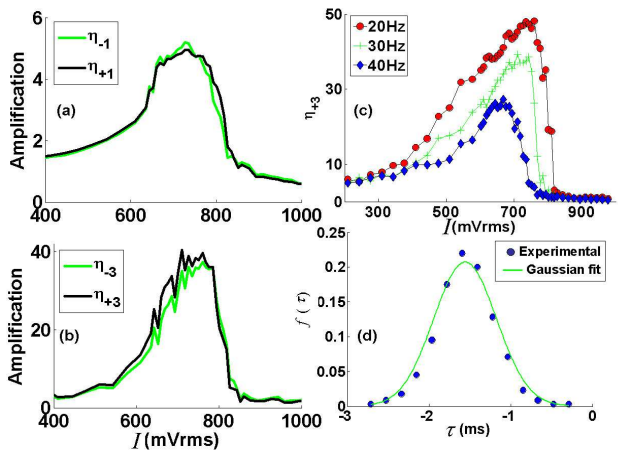}%
\caption{(Color online). (a),(b) Spectral amplification $\eta_{k}(D)$
($k=\pm1$ and $k=\pm3)$ vs. voltage noise intensity. (c) The spectral
amplification $\eta_{+3}(D)$ vs. noise intensity for three AM frequencies. (d)
Measurement of the probability density $f(\tau),$ where $\tau$ is the
difference between the time of the transition $S_{l}\rightarrow S_{h}$ and the
time at which the modulation amplitude gets its maximal value.}%
\end{center}
\end{figure}

The dependence of the spectral amplification $\eta_{3}(D)$ on $I$ for three AM
frequencies $\Omega=20%
\operatorname{Hz}%
$ ,$30%
\operatorname{Hz}%
$, and $40%
\operatorname{Hz}%
$ is shown in Fig. 4c for $A_{\operatorname{mod}}=10\%$. As predicted
theoretically,\cite{Jung} amplification \ is monotonically decreasing with
$\Omega$.

Finally, we demonstrate the method proposed in\cite{BaleeghEscRate} for
extracting transition rates from SR measurements. Near the maximum (minimum)
points of the AM signal, the rate $\Gamma_{1}$ $(\Gamma_{2})$ obtains its
largest value, which is denoted by $\Gamma_{m1}$ $(\Gamma_{m2}).$ Let $\tau$
be the difference between the time of the transition $S_{l}\rightarrow S_{h}$
and the time at which the modulation amplitude gets its maximal value (namely,
the time at which $\Gamma_{1}=\Gamma_{m1}).$ The probability density of the
random variable $\tau$, which is denoted by $f(\tau)$, is experimentally
derived from 1000 modulation cycles sampled in the time domain (see Fig. 4d).
The solid line represents a Gaussian function fitted to the measured
probability density. The rate $\Gamma_{m1}$ can be estimated from the
expectation value $\mu_{\tau}$ and the variance $\sigma_{\tau}^{2}$ of $\tau$
\cite{BaleeghEscRate} by $\Gamma_{m1}=-\mu_{\tau}/\sigma_{\tau}^{2}$, yielding
$\Gamma_{m1}=10.5%
\operatorname{kHz}%
$.

In conclusion, stochastic resonance has been demonstrated in a nanomechanical
resonator. The resonator was tuned to its bistability region by an intense
pump near a point of equal transition rates between its states. An amplitude
modulation is used to modulate the activation barrier between the states. When
noise is injected, the resonator's response exhibits noise dependent
amplification. We measure the resonator's displacement in the time and
frequency domains, the spectral amplification and statistics of the jumps
time. SR could be very useful in nanomechanical devices as a mean to implement
on-chip mechanical amplification and to increase the signal to noise ratio.

This work is supported by the Israeli ministry of science, Intel Corp.,
Israel-US binational science foundation, The Russell Berrie Nanotechnology
Institute , and by Henry Gutwirth foundation.

\end{document}